\documentclass[prl,twocolumn,aps,10pt,superscriptaddress,notitlepage]{revtex4-1}  
\usepackage{epsfig,amsmath,amssymb,amsfonts,graphicx,color,calc,epstopdf,stmaryrd}

\let\a=\alpha \let\b=\beta \let\g=\gamma 
   
\let\l=\lambda    
\let\s=\sigma \let\t=\tau \let\f=\varphi 
   
\let\D=\Delta   
\let\Si=\Sigma   
 \let\r=\rho  \let\io=\infty

\def\to{\rightarrow} \def\la{\left\langle} \def\ra{\right\rangle}

\def\wh{\widehat}

\def\de{\mathrm d}

\newcommand{\beq}{\begin{equation}} 
\newcommand{\eeq}{\end{equation}}
\newcommand{\ba}{\begin{eqnarray}}
\newcommand{\ea}{\end{eqnarray}}

\begin{document}

\title{
Shear yielding and shear jamming of dense hard sphere glasses
}

\author{Pierfrancesco Urbani}
\affiliation{Institut de physique th\'eorique, Universit\'e Paris Saclay, CNRS, CEA, F-91191 Gif-sur-Yvette}

 \author{Francesco Zamponi}
 \affiliation{Laboratoire de Physique Th\'eorique, ENS \& PSL University, UPMC \& Sorbonne Universit\'es, 
UMR 8549 CNRS, 75005 Paris, France}

\begin{abstract}
We investigate the response of dense hard sphere glasses to a shear strain, in a wide range of pressures ranging from
the glass transition to the infinite-pressure jamming point. The phase diagram in the density-shear strain plane is calculated
analytically using the mean field infinite dimensional solution.
We find that just above the glass transition, the glass generically yields at a finite
shear strain. The yielding transition, in the mean field picture, is a spinodal point in presence of disorder. At higher densities,
instead, we find that the glass generically jams at a finite shear strain: the jamming transition prevents yielding. The shear yielding
and shear jamming lines merge in a critical point, close to which the system yields at extremely large shear stress. Around this
point, a highly non-trivial yielding dynamics characterized by system-spanning disordered fractures is expected.
\end{abstract}

\maketitle

\paragraph*{Introduction --}
The response of glasses to a shear strain is extremely complex and has always been the subject of much interest,
for fundamental and technological reasons~\cite{AK79,FL98,SWS07,BL11,RTV11}. 
While at small enough strains the solid responds
elastically, at moderate strains the response is characterized by small intermittent drops of the shear stress. At larger
strains, the stress drops abruptly when the glass yields. Above yielding, the stress remains approximately constant upon
increasing strain, and the system flows~\cite{BL11,RTV11}. For soft interaction potentials, it has been established that
both low-stress intermittency and large-stress flow are due to ``plastic'' events at which small regions of the material
-- called ``shear transformations'' --
fail under stress~\cite{FL98,BL11,PRB16,DHPS16}. 
The energy relaxed by the failure is propagated elastically 
through the system, leading to failure in
other regions. 
The stress-strain curves can be well described in the flow regime
by elasto-plastic models, that describe mesoscopically the coupling between
failing plastic regions~\cite{SLHC97,HL98,ABMB15,LW16},
and the plastic regions themselves
have been identified quite precisely in numerical simulations~\cite{PRB16,DHPS16}.

The situation is quite different for dense hard sphere glasses, that are good models of colloidal and granular glasses.
These solids, due to the hard sphere constraints, are characterized
by a critical ``random close packing'' or ``jamming'' density at which a rigid isostatic network of particle contacts emerges, 
inducing a divergence of the pressure~\cite{OLLN02,PZ10,TS10}.
Around the jamming point, due to the emergent contact network, perturbing a particle leads to a 
macroscopic rearrangement of the whole solid~\cite{Wy12,LDW13,DLBW14}:
continuum elasticity breaks down~\cite{CR00, LDDW14, BU16, FS16} 
and solid dynamics is characterized by system-spanning avalanches during which the system relaxes along
strongly delocalized soft modes~\cite{WSNW05,BW09b, FS16}. Clearly, in this regime the ``shear transformations'' picture becomes inappropriate.

The aim of this Letter is to characterize
the response of a dense hard sphere glass to a static shear strain (i.e. in the regime where the solid responds by a static stress, without flowing), 
all the way from the glass transition to the jamming
regime, within a mean field approach.
We find that at lower densities, slightly above the glass transition, 
the hard sphere glass responds in a way similar to soft particle glasses:
an elastic regime is followed by an intermittent regime before the system yields (``shear yielding''). At larger densities, close to jamming, 
the situation is radically different. 
Before yielding,
a jamming transition happens due to shear: at the transition a rigid network of contacts is formed and the pressure diverges (``shear jamming''). 
The shear jamming transition is in the same universality class of the jamming transition
at zero shear~\cite{LDW12,BGLNS16}, 
and it is characterised by non-trivial critical exponents that appear in the interparticle force and gap distributions~\cite{Wy12,nature,CCPZ12}. 
Most importantly,
the shear yielding and shear jamming lines merge in a critical point. 
Around this point, because the system yields at extremely large (diverging) pressure and shear stress, in a regime of incipient jamming, we expect
a highly non-trivial yielding dynamics, characterized by system-spanning disordered fractures.

\paragraph*{Glass preparation protocol --} 
We consider a system of $N$ identical $d$-dimensional hard spheres, in the thermodynamic limit 
at constant number density $\r$ and volume fraction $\f$. 
We consider the limit $d\to\io$, with constant $\wh\f = 2^d \f / d$,
in which the liquid and glass properties can be computed
exactly within the mean field Random First Order Transition scenario~\cite{KW87,CKPUZ16}.
For hard spheres, the infinite dimensional limit usually provides qualitatively good predictions for the phase diagram
of low-dimensional systems~\cite{CKPUZ16}, especially around jamming~\cite{nature}, 
and finite-dimensional effects can be studied through numerical simulations~\cite{CCPZ12,CJPZ14}.
Also, for $d>3$ polydispersity is not needed, as monodisperse hard spheres are a very good glass-forming system~\cite{SDST06,CIMM10}.

\begin{figure}[t]
\includegraphics[width=\columnwidth]{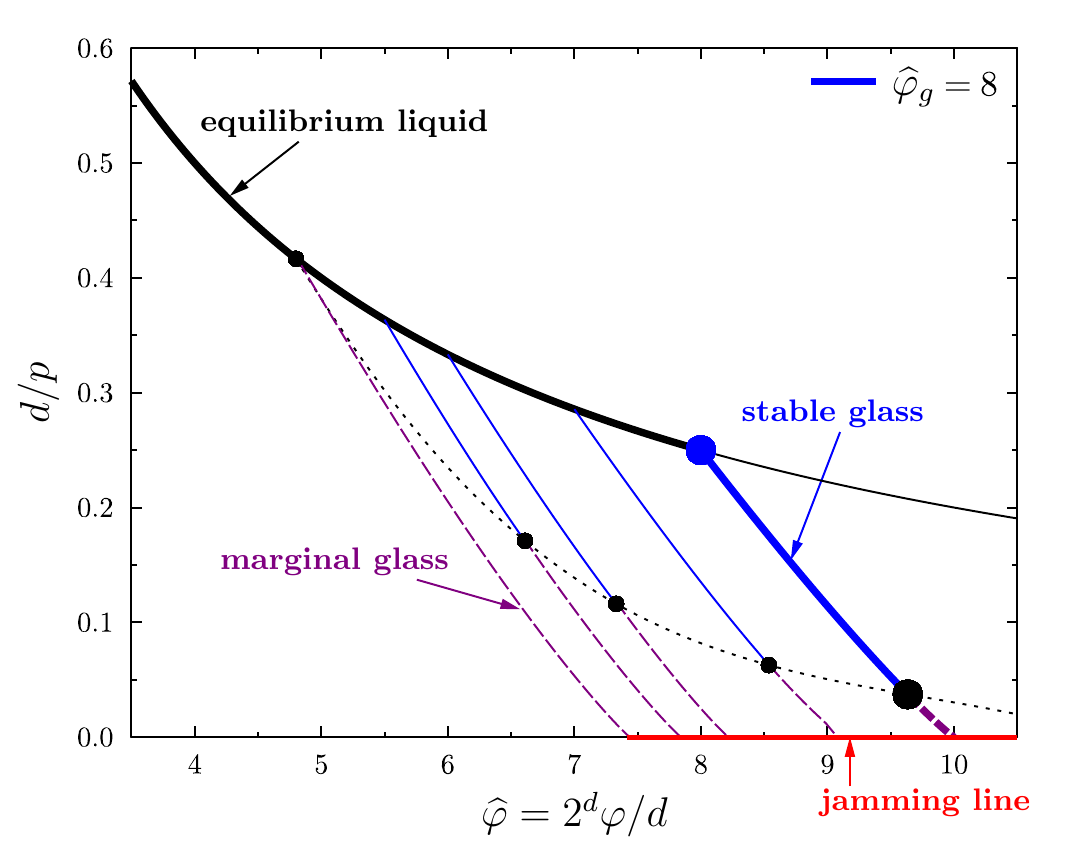}
\caption{Inverse reduced pressure $d/p$
versus packing fraction $\wh\f = 2^d \f/d$ (both scaled to remain finite for $d\to\io$)
during a slow compression~\cite{RUYZ15,RU16}.
The liquid EOS is $d/p = 2/\wh\f$. The dynamical transition
$\wh\f_{\rm d}$ is marked by a black dot. 
We focus on a liquid slowly compressed up to $\wh\f_{\rm g}=8$ (blue full circle).
From that point on, the system is followed in a restricted equilibrium confined to the glass state (full blue line).
At high pressure, the glass state becomes marginally stable. Jamming is reached around $\wh\f_{\rm j} \approx 10$.
The thick lines indicate the specific glass we follow in this paper. Other glasses corresponding
to different $\wh\f_{\rm g}$ (different compression rates) are plotted with thinner lines.
}
\vskip-5pt
\label{fig:compression}
\end{figure}

During a slow cooling of a liquid, the relaxation
time scale $\t_\a(\wh\f)$ becomes extremely large around the Mode-Coupling density $\wh\f_{\rm d}$, 
but one can still equilibrate up to quite larger values of $\wh\f$, either by brute force~\cite{BEPPSBC08} 
or by means of smart numerical algorithms~\cite{GP01,SEP13,BCNO15,BCJPSZ16} 
and smart experimental protocols~\cite{Sw07}. 
Once equilibration at some $\wh\f_{\rm g}>\wh\f_{\rm d}$ has been achieved, one can 
focus on time scales $\t_{\rm exp} \ll \t_{\rm \a}(\wh\f_{\rm g})$, in such a way that the system 
remains confined in the glass state selected in equilibrium at $\wh\f_{\rm g}$. 

In the mean field limit $d\to\io$, the liquid relaxation time diverges above $\wh\f_{\rm d}$,
and the dynamics is completely arrested~\cite{KW87,Go99,MKZ16}. 
The separation of time scales thus becomes very sharp
as $\t_{\rm exp} \ll \t_{\rm \a}(\wh\f_{\rm g}) \to \io$.
The ``state following'' formalism is designed to describe this regime~\cite{FP95,KZ10,KZ13},
in which a typical equilibrium configuration selects a long-lived glass basin
which is then adiabatically followed upon increasing
the density $\wh\f \geq \wh\f_{\rm g}$ and applying a shear strain $\g$~\cite{RUYZ15,RU16,CKPUZ16}.
In particular, the method gives the reduced pressure $p = \b P/\r$ and shear stress $\s = \b \Si$ of the glass.

The pressure-density equation of state in absence of shear has been studied in~\cite{RUYZ15,RU16}
(Fig.~\ref{fig:compression}). We focus on a liquid compression that remains in equilibrium until 
$\wh\f_{\rm g}=8 > \wh\f_{\rm d} \approx 4.8$ (this is representative of a typical situation), 
and we follow the corresponding glass in a restricted equilibrium. 
This glass undergoes
a Gardner phase transition to a marginally stable state~\cite{nature,RUYZ15,RU16,CKPUZ16}, and then jams at a density $\wh\f_{\rm j} \approx 10$.
The phase diagram of Fig.~\ref{fig:compression} qualitatively agrees with $3d$ numerical simulations~\cite{BCJPSZ16}.

\paragraph*{Stress-strain curves --} 
The glass prepared at $\wh\f_{\rm g}$ is first adiabatically compressed to $\wh\f > \wh\f_{\rm g}$, and then
a shear strain $\g$ is applied.
At the {\it replica symmetric} level~\cite{MPV87}, 
the glass free energy $f_{\rm g}(\g,\wh\f; \D,\D_{\rm r})$
can be exactly computed in $d\to\io$ as a function of two order parameters~\cite{RUYZ15}:
$\D$ is the mean square displacement (MSD) in the glass state at $(\wh\f,\g)$,  and $\D_{\rm r}$
is the relative MSD of a typical configuration of the glass at $(\wh \f_{\mathrm g},\g=0)$ and a typical configuration 
of the same glass once followed up to $(\wh\f,\g)$ (see~\cite{RUYZ15} for the precise mathematical definition). 
Both are obtained by setting the derivatives of $f_{\rm g}$ to zero. Once $\D,\D_{\rm r}$ are determined,
the average reduced pressure $p$ and average
stress $\s$ are derivatives of $f_{\rm g}$ with respect to $\wh\f$ and $\g$, respectively.

\begin{figure*}[t]
\includegraphics[width=.95\textwidth]{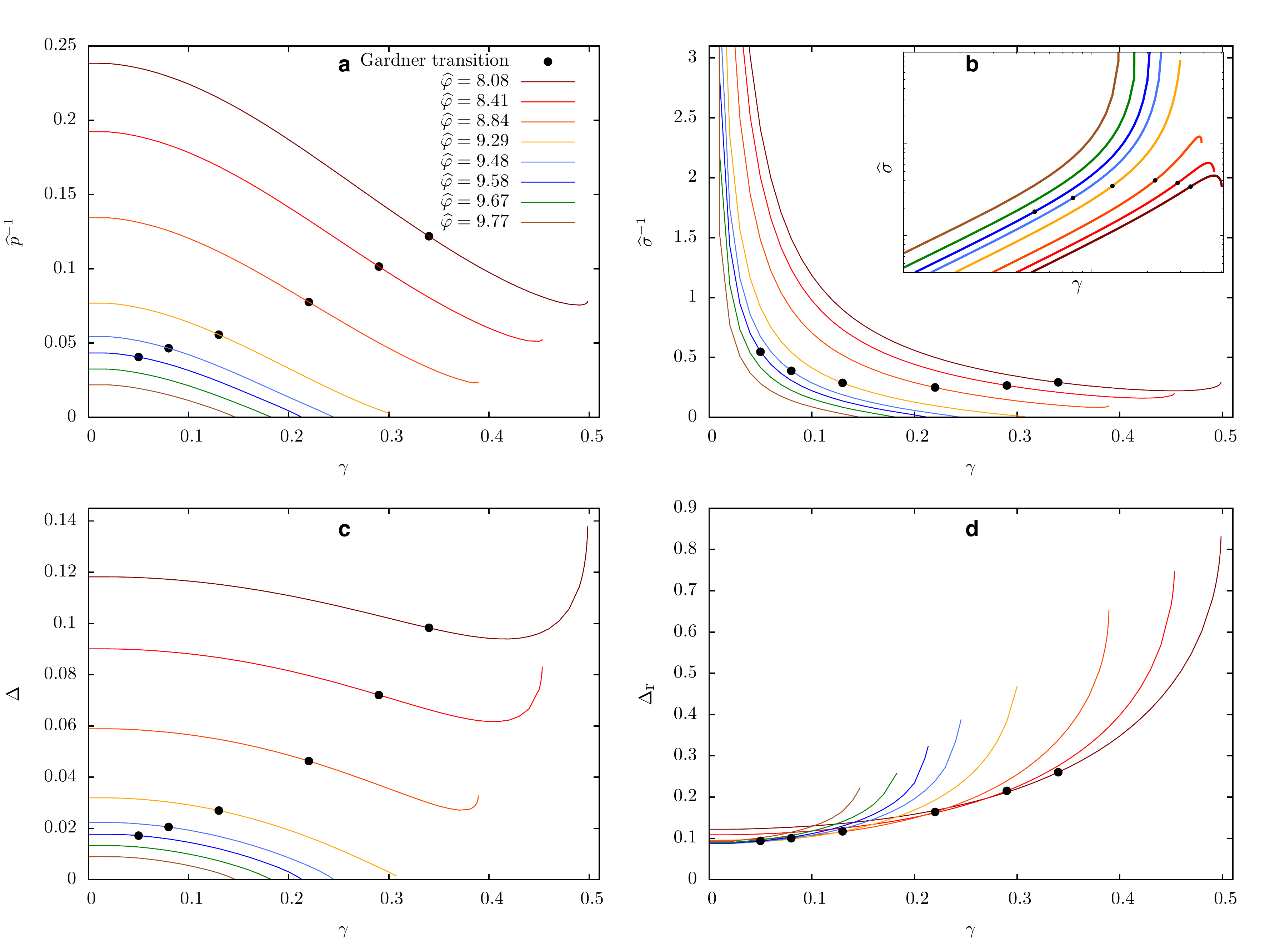}
\caption{
Applying adiabatically a shear strain $\g$ on a glass prepared at equilibrium at $\wh\f_{\rm g}=8$ and adiabatically compressed to 
$\wh\f \in [\wh\f_{\rm g}, \wh\f_{\rm j}]$. 
The black dots along the lines represent the Gardner transition.
{\bf (a)} Inverse reduced pressure $d/p\equiv \wh p^{-1}$ vs $\g$. At lower $\wh\f$, the pressure is finite until
the system yields at $\g_{\rm y}(\wh\f)$. At higher density, the pressure diverges at the shear jamming point $\g_{\rm j}(\wh\f)$.
{\bf (b)} Inverse of the reduced shear stress $\wh \s^{-1}\equiv d/\sigma$ vs $\g$. The behavior is very similar to the pressure. At lower $\wh\f$, the stress overshoots before yielding. 
At higher $\wh\f$, the stress diverges at $\g_{\rm j}$ without any overshoot. 
The inset shows the behavior of $\wh \s$ vs $\g$ in log scale.
{\bf (c)} The glass MSD $\D$ vs $\g$. At lower $\wh\f$, $\D$ remains finite at yielding. At higher $\wh\f$, $\D$ vanishes
at shear jamming. {\bf (d)} The MSD $\D_{\rm r}$ between the initial equilibrium configuration at $\wh\f_{\rm g}$ and the one
at $(\wh\f,\g)$. At lower $\wh\f$, $\D_{\rm r}$ remains finite and displays a square-root singularity at yielding, such that $\de\D_{\rm r}/\de\g \to \io$ for $\g\to\g_{\rm y}$.
At higher density, $\D_{\rm r}$ remains finite at shear jamming with no singularity. \newline
}
\vskip-10pt
\label{fig:strain}
\end{figure*}

All the four quantities $p$, $\s$, $\D$ and $\D_{\rm r}$ are reported in Fig.~\ref{fig:strain} as functions of $\g$ for several values of $\wh\f$. We observe
a different behavior at lower densities close to $\wh\f_{\rm g} =8$ and at higher densities close to $\wh\f_{\rm j}\approx 10$. For lower $\wh\f$,
there is first a linear elastic regime $\s\sim \mu \g$, followed by a stress overshoot before the system finally yields at $\g_{\rm y}(\wh\f)$. 
At the mean field, replica symmetric level, the yielding point is defined by the fact that
stress, pressure, $\D$ and $\D_{\rm r}$ display a square-root singularity, e.g. $p - p_{\rm y}(\wh\f) \propto \sqrt{\g_{\rm y}(\wh\f) - \g}$,
because yielding is akin to a spinodal: the solution of the stationarity equations for $\D,\D_{\rm r}$ merges
with another unphysical solution and disappears in a bifurcation-like manner. 
Equivalently, the square-root singularity is due
to the vanishing of a {\it longitudinal mode} $\l_{\rm L} \propto \de^2 f_{\rm g}/\de \D_{\rm r}^2$ at $\g_{\rm y}(\wh\f)$. 
It also implies that
there is a diverging susceptibility at $\g_{\rm y}(\wh\f)$, related to the fluctuations of $\D_{\rm r}$:
$\chi_{\rm L} \sim \la \D_{\rm r}^2 \ra - \la \D_{\rm r} \ra^2 \propto 1/\l_{\rm L}$.
For higher $\wh\f$, instead, we observe that pressure and stress increase fast
and both diverge at a shear jamming point $\g_{\rm j}(\wh\f)$, where $\D \to 0$ and $\D_{\rm r}$ remain finite. 

\begin{figure}[t]
\includegraphics[width=\columnwidth]{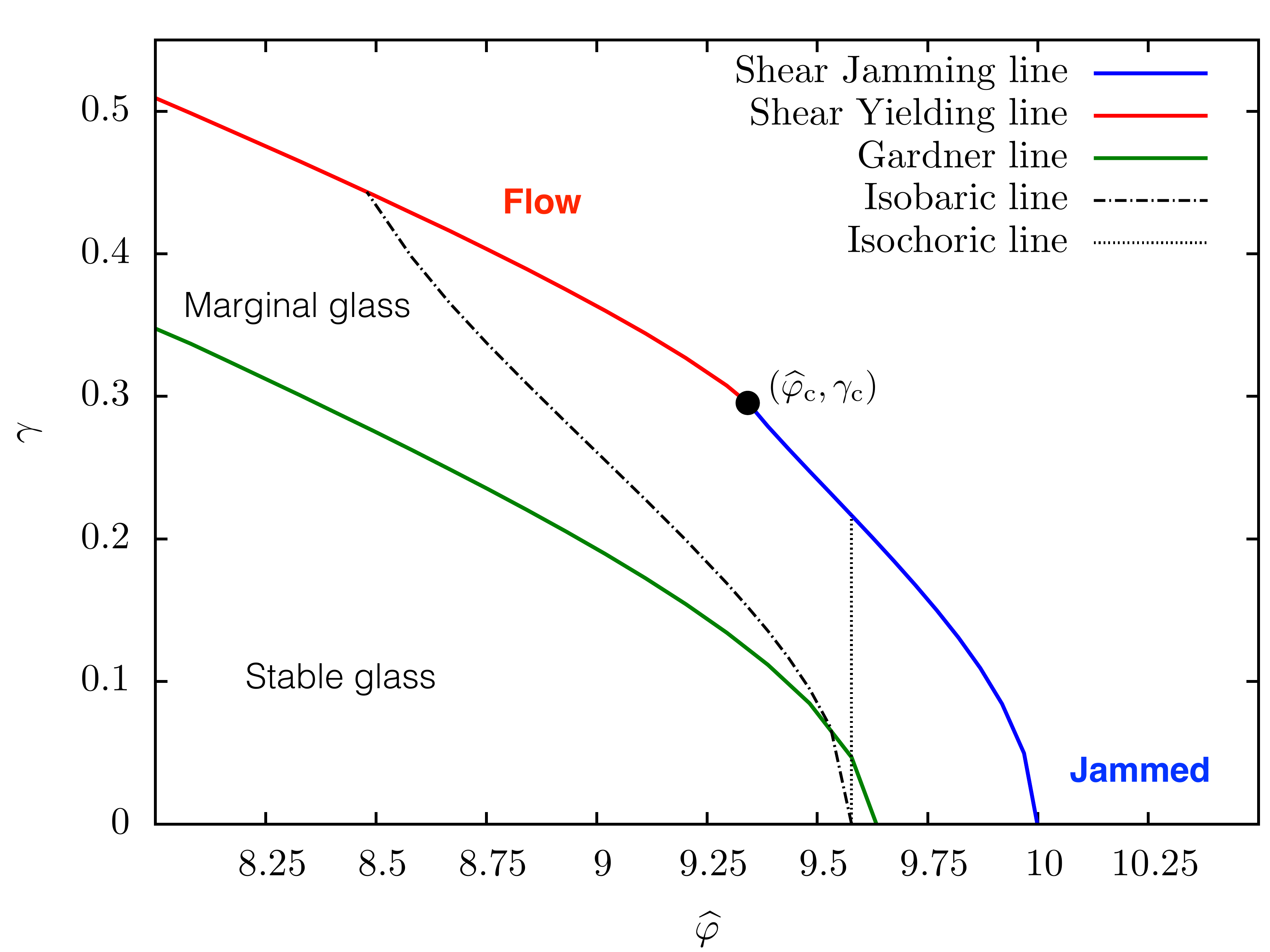}
\vskip-10pt
\caption{
Phase diagram of the glass prepared in equilibrium at $\wh\f_{\rm g}=8$, and followed adiabatically at density $\wh\f > \wh\f_{\rm g}$ and shear strain $\g$.
The shear jamming line $\g_{\rm j}(\wh\f)$ and the shear yielding line $\g_{\rm y}(\wh\f)$ are plotted. The two lines merge at a critical point
$(\wh\f_{\rm c}, \g_{\rm c})$. At this special point, yielding happens at infinite pressure/strain. For $\wh\f \lesssim \wh\f_{\rm c}$, 
yielding happens at $\g \sim \g_{\rm c}$ with extremely large pressure/strain.
The vertical dotted line is the isochoric line of a glass prepared at $\wh \f>\wh \f_{\textrm{c}}$ and then strained at fixed volume until it shear jams.
The dotted-dashed line represent the isobaric line of a glass prepared at the same initial packing fraction but then strained at fixed pressure. The shear jamming transition is thus avoided and the glass yields for sufficiently high strains.
}
\vskip-15pt
\label{fig:PD}
\end{figure}

\paragraph*{Phase diagram --}
In Fig.~\ref{fig:PD} the shear yielding line $\g_{\rm y}(\wh\f)$ and the shear jamming line $\g_{\rm j}(\wh\f)$ are reported in the $(\wh\f,\g)$ plane
for $\wh\f_{\rm g}=8$. We observe a re-entrant shear jamming line, moving to lower densities for increasing $\g$. The shear yielding 
line $\g_{\rm y}$ decreases upon increasing $\wh\f$. It is possible to show analytically that the two lines merge at a critical point
$(\wh\f_{\rm c}, \g_{\rm c})$, at which the system is both jammed (because $\D=0$, $p=\io$, $\s=\io$) and yielding, because the longitudinal 
mode vanishes indicating an instability of $\D_{\rm r}$ (which remains finite at the critical point, but has infinite derivative).
Note that beyond the yielding point, the solid phase is unstable and the systems starts to flow: a fixed stress $\s$
corresponds to a finite {\it shear rate} $\dot\gamma$. In this regime, the state following formalism is not appropriate (both $\D_{\rm r}$ and $\D$
are formally infinite) and a fully dynamical treatment is needed, which goes beyond the scope of this work.
We also computed the phase diagram for different values of $\wh\f_{\rm g}$ (not shown). We find that the critical density $\wh\f_{\rm c}$ 
moves towards $\wh\f_{\rm j}$ upon decreasing $\wh\f_{\rm g}$, which implies that the shear jamming line shrinks and eventually disappears for 
poorly equilibrated glassy states with $\wh\f_{\rm g} \approx \wh\f_{\rm d}$.

\paragraph*{Marginal stability --}
As previously found in~\cite{RUYZ15,RU16}, the replica symmetric solution used to compute the results of Fig.~\ref{fig:strain}
becomes unstable in a region of the phase diagram delimited by the {\it Gardner transition line} $\g_{\rm G}(\wh\f)$. Beyond this line, the order
parameter $\D$ becomes a function $\D(x)$ defined for $x\in[0,1]$ and the glass free energy is a functional $f_{\rm g}[\g,\wh\f; \D(x),\D_{\rm r}]$.
The resulting {\it full replica symmetry breaking} solution~\cite{MPV87} is characterized by marginal stability: one of the derivatives
of the free energy (the {\it replicon} mode) is identically vanishing in the marginally stable phase, 
leading to a diverging susceptibility and the breakdown of standard elasticity~\cite{CKPUZ16, BU16}.
The function $\D(x)$ and $\D_{\rm r}$ are determined by setting the (functional) derivatives of $f_{\rm g}$ to zero.
Although we did not solve the resulting equations numerically 
(see~\cite{RU16} for a computation of the stress-strain curves at $\wh\f_{\rm g})$, the phase diagram
remains qualitatively similar to Fig.~\ref{fig:PD}. Indeed, we can show analytically that 
{\it (i)}~shear jamming is characterized by the vanishing of $\D(1)$, the self MSD of the 
glass states, which induces a divergence of pressure $p$ and stress $\s$.
On the shear jamming line, the critical properties are the same
of the jamming point at zero strain~\cite{nature}: the inter-particle force and gap distributions display 
power-law behavior, with non-trivial exponents
that are constant along the shear jamming line. {\it (ii)}~Shear yielding is still characterized by the vanishing of $\l_{\rm L} \propto \de^2 f_{\rm g}/\de \D_{\rm r}^2$, 
which induces a divergence
of the fluctuations of $\D_{\rm r}$. However, the critical properties on the shear yielding line remain to be understood.
{\it (iii)}~The two lines merge at a critical point where both $\D(1)=0$ and
$\l_{\rm L}=0$. 

\paragraph*{Comparison with numerics and experiments --} 
Many experimental and numerical works have studied both shear yielding and shear jamming. 
In particular, simulations of athermal systems~\cite{BGLNS16,HB09,HCB10,PCC11,Seto13,VS16}
and experiments on granular materials~\cite{CD09,Bi11,Fall08} found 
a re-entrant shear jamming line. 

The phase diagram in our Fig.~\ref{fig:PD} holds for a specific protocol: a thermal system that is prepared in a well equilibrated initial 
state ($\wh\f_{\rm g} > \wh\f_{\rm d}$),
to which compression and shear strain are applied. 
In a systematic study of a (frictionless) athermal system~\cite{BGLNS16}, 
it has been found that
the re-entrance of the shear jamming line is a finite size effect and
disappears when $N\to\io$.
There are two possible explanations for this difference. First, it could be due to the lack of 
initial equilibration of the samples used in~\cite{BGLNS16}, which are prepared by quenching instantaneously from
infinite temperature: this is consistent with our finding that poorly equilibrated thermal systems do not display shear 
jamming.
Also,
while for thermal hard spheres (any $T>0$) entropic forces
stabilize the solid phase in the region delimited by the shear yielding
and shear jamming lines in Fig.~\ref{fig:PD},
athermal systems ($T=0$) below jamming are not rigid (at least for small $\g$) because
both the pressure $P = T p$ and the stress $\Si = T \s$ vanish identically at $T=0$.
It is thus possible that they have a very different dynamics upon
application of shear strain~\cite{IBS12}, in which case it would be difficult to compare athermal system with our theory.
Additional numerical simulations are needed to clarify this issue.

Granular materials under
tapping could instead be equivalent to thermal systems and display a re-entrance that persists for $N\to\io$, but a finite-size
study has not been performed in this case~\cite{CD09,Bi11,Fall08}.
Also, the results of~\cite{JPRS16,TB16} on shear yielding support the idea that this transition is similar to a spinodal point in presence of disorder.
A more direct comparison can be made between our theory and very recent simulations of thermal hard spheres under shear~\cite{JY16}.
Our predictions are qualitatively compatible with these numerical results. 
However, none of these studies has investigated
the coalescence of the shear yielding and shear jamming lines at  $(\wh\f_{\rm c}, \g_{\rm c})$, and the plastic dynamics around the critical point, 
which is the most interesting result of this work.

\paragraph*{Conclusions --}
We investigated the phase diagram of a dense hard sphere glass, prepared in equilibrium at $\wh\f_{\rm g} > \wh\f_{\rm d}$, 
and followed adiabatically to density $\wh\f$ and shear strain $\g$. The phase diagram in the $(\wh\f, \g)$ plane (Fig.~\ref{fig:PD}) generically
displays a shear yielding line when $\wh\f \gtrsim \wh\f_{\rm g}$, and a shear jamming line when $\wh\f \lesssim \wh\f_{\rm j}$. The two
lines merge at a critical point, around which the system yields at extremely large pressure and shear stress.

Although our results are derived in a mean field setting, we expect that they describe accurately 
the critical exponents associated to the shear jamming line in finite dimensions, 
as it is the case for $\g=0$~\cite{nature,LDW12}. 
Indeed, the shear jamming line has the same critical properties of the isotropic jamming transition~\cite{BGLNS16}.
On the contrary, even at the mean field level, 
the critical properties of the shear yielding line are not fully understood, because this line falls in a region where the glass is marginally stable
and a full replica symmetry breaking scheme is needed~\cite{RU16}. Moreover,
because the yielding transition is a spinodal point in presence of disorder, 
it cannot be strictly described by mean field in any dimension~\cite{FPRR11,BNT16}.
A detailed characterization of the yielding transition is thus a very difficult task and it is certainly a
very important line for future research. 
However, at the critical point $(\wh \f_{\mathrm c},\g_{\mathrm c})$, we conjecture that the system sizes where finite $d$ corrections become important diverge, 
so that the mean field theory of the yielding transition can likely become exact close to $(\wh \f_{\mathrm c},\g_{\mathrm c})$.
The plastic dynamics
around $(\wh \f_{\mathrm c},\g_{\mathrm c})$ is expected to be strongly different from the one of soft glasses. Its analytical and numerical investigation is another 
very interesting subject for future work. 
Systematic numerical~\cite{JY16} and experimental~\cite{SD16} investigation of the phase diagram in Fig.~\ref{fig:PD} will be 
of great help to fully understand the interplay of yielding and jamming in amorphous solids.

\paragraph*{Acknowledgments --}
We warmly thank E.~Agoritsas, M. Baity Jesi, J.-L.~Barrat, G.~Biroli, O.~Dauchot, E.~DeGiuli, Y.~Jin, C.~Rainone, S.~Sastry, M.~Wyart, and H.~Yoshino
for many useful discussions.
This work was supported by a grant from the Simons Foundation (\#454955, Francesco Zamponi).


\end{document}